\documentclass[10 pt, conference]{ieeeconf}

\IEEEoverridecommandlockouts 
\makeatletter

\let\proof\@undefined
\let\endproof\@undefined
\makeatother
\let\labelindent\relax 

    \usepackage{jhoagg}
    \usepackage{mathtools}
    \usepackage{amsfonts}
    \usepackage{amssymb,latexsym}
    \usepackage[psamsfonts]{eucal}
    \usepackage{amsthm}
    \usepackage{graphicx}
    \usepackage[inline]{enumitem}
    \usepackage{indentfirst}
    \usepackage{setspace}
    \usepackage{microtype}
    \usepackage{threeparttable}
    \usepackage{float}
    \usepackage{cleveref}
    \usepackage{afterpage}
    \usepackage{placeins}
    \usepackage{cancel}
    \usepackage{leftidx}
    \usepackage{breqn}
    \usepackage[export]{adjustbox}
    \usepackage{comment}
    \usepackage[ruled, linesnumbered, lined]{algorithm2e}
    \usepackage[dvipsnames]{xcolor}
    \usepackage{xcolor}
    \usepackage[most]{tcolorbox}
    \usepackage{soul}
    \usepackage[noadjust]{cite}
    \usepackage{standalone}
    \usepackage{url}

    \usepackage{pgf}
    \usepackage{pgfplots}
    \usepgflibrary{plothandlers}
    \usepgfplotslibrary{external} 
    \pgfkeys{/pgf/number format/.cd,1000 sep={}}  
    \pgfplotsset{compat=1.18}

    \let\originalleft\left
    \let\originalright\right
    \renewcommand{\left}{\mathopen{}\mathclose\bgroup\originalleft}
    \renewcommand{\right}{\aftergroup\egroup\originalright}
    
    \usepackage{fancyhdr}
    \fancypagestyle{firststyle}
    {\fancyfoot{} 
    \fancyfoot[L]{\textit{Accepted for the 2026 American Control Conference (ACC)}}
     
    }
    
    \newtheoremstyle{indentedplain}
      {3pt}      
      {3pt}      
      {\rmfamily}
      {1em}      
      {\bfseries}
      {.}        
      {0.5em}    
      {}         
    
    \newtheoremstyle{indentedremark}%
      {3pt}{3pt}{}
      {1em}{\bfseries}{.}{0.5em}{}

    \theoremstyle{indentedplain}
    \newtheorem{theorem}{Theorem}
    
    \newtheorem{proposition}{Proposition}
    \newtheorem{lemma}{Lemma}
    
    \newtheorem{definition}{Definition}
    \newtheorem{fact}{Fact}

    \theoremstyle{indentedremark}

    \allowdisplaybreaks

    \newlist{enumA}{enumerate}{1}
    \setlist[enumA,1]{label=(A\arabic*),leftmargin=1cm}

    \newlist{enumC}{enumerate}{1}
    \setlist[enumC,1]{label=(C\arabic*),leftmargin=1cm}

    \newlist{enumO}{enumerate}{1}
    \setlist[enumO,1]{label=(O\arabic*),leftmargin=1cm}

    \usepackage{cite}
    \usepackage{accents}
    
    \newlength\figureheight 
    \newlength\figurewidth

    \allowdisplaybreaks
    
    \graphicspath{ {Figures/} }
    
    \DeclareMathAlphabet{\mathcal}{OMS}{cmsy}{m}{n} 

    \crefname{equation}{}{}
    \crefname{figure}{Fig.}{Figure}
    
    \usepackage[ruled, linesnumbered, lined]{algorithm2e}
    \SetKwInput{KwInit}{Initialize}
    \SetKwFunction{SafeSet}{SafeSet}
    \SetKwFunction{GetSafe}{GetSafe}
    \SetKwFunction{GetUnsafe}{GetUnsafe}
    
    \SetCommentSty{mycommfont}
    \newcommand{\ubar}[1]{\underaccent{\bar}{#1}}
    \newlist{enumalph}{enumerate}{1}
    \setlist[enumalph]{label=\textit{(\alph*)}}

\begin{document}

\title{Safe Landing on Small Celestial Bodies with Gravitational Uncertainty Using Disturbance Estimation and Control Barrier Functions}
\author{Felipe Arenas-Uribe, T. Michael Seigler, and Jesse B. Hoagg
\thanks{F. Arenas-Uribe, T. M. Seigler, and Jesse B. Hoagg are with the Department of Mechanical and Aerospace Engineering, University of Kentucky, Lexington, KY, USA. (e-mail: felipearur@uky.edu, tmseigler@uky.edu, jesse.hoagg@uky.edu).}
}                                                                                                                                                                                                                                                                             
\maketitle

\thispagestyle{firststyle}

\begin{abstract}
    Soft landing on small celestial bodies (SCBs) poses unique challenges, as gravitational models poorly characterize the higher-order gravitational effects of SCBs. Existing control approaches lack guarantees for safety under gravitational uncertainty. This paper proposes a three-stage control architecture that combines disturbance estimation, trajectory tracking, and safety enforcement. An extended high-gain observer estimates gravitational disturbances online, a feedback-linearizing controller tracks a reference trajectory, and a minimum-intervention quadratic program enforces state and input constraints while remaining close to the nominal control. The proposed approach enables aggressive yet safe maneuvers despite gravitational uncertainty. Numerical simulations demonstrate the effectiveness of the controller in achieving soft-landing on irregularly shaped SCBs, highlighting its potential for autonomous SCB missions.
\end{abstract}

\section{Introduction}

Small celestial body (SCB) missions have attracted significant attention, both for their scientific value and their relevance to planetary defense. Sample return missions have provided insights into the formation and composition of celestial bodies, while near-Earth objects also drives active monitoring research. Notable examples include NASA’s OSIRIS-REx, JAXA’s Hayabusa and ESA’s Rosetta missions. Despite these advances, fully autonomous soft-landing on a SCB remains challenging due to uncertainties in gravitational models and poorly mapped environments.

Uncertainties in gravitational field models stem primarily from estimating SCB mass, which typically carries a 10--15\% measurement error \cite{Bull2021Mass_measurement}. Beyond mass, gravitational field models must also capture higher-order gravitational effects arising from an SCB’s non-uniform density distribution. To account for these effects, higher-order models are commonly used \cite{scheeres2012orbital,arenasuribe2026higherordergravitationalmodelstutorial}. However, accurately determining the model parameters requires high-resolution remote sensing or in-situ measurements that are difficult to acquire \cite{Penarroya2022}.

To meet safety requirements, guidance and control frameworks typically rely on trajectory generation algorithms executed offline and an online feedback tracking control law \cite{malyuta_advances_2021}. For a review of SCB soft-landing trajectory generation approaches see \cite[Section 3.3]{malyuta_advances_2021}. Feedback tracking controllers for these maneuvers have employed robust approaches, such as $H_\infty$ control \cite{simplicio_synthesis_2018} and adaptive control \cite{tiwari_direct_2023}. Linear extended high-gain observers (EHGO) have also been applied to estimate and compensate gravitational uncertainties \cite{dunham_constrained_2016}. While these approaches provide robustness to disturbances, they offer no guarantees on satisfying safety constraints. Model Predictive Control (MPC) methods explicitly incorporate state and input constraints \cite{sanchez_predictive_2018, tiwari_direct-adaptive_2022, van_leeuwen_nonlinear_2022}. However, such formulations rely on approximations of the nonlinear dynamics of the system. As a result, constraint satisfaction is not guaranteed unless additional constraint-tightening mechanisms are introduced, which makes the algorithm computationally expensive.

These limitations motivates the use of control barrier functions (CBFs), which have been used as a mechanism for safety constraint satisfaction \cite{ames_control_2019}. CBFs have been applied in spacecraft maneuvers such as satellite orbit transfers \cite{ong_intermittent_2023}, proximity maneuvers \cite{breeden_robust_2023} and formation flying \cite{kamat_electromagnetic_2026}. However, their application into SCB soft-landing remains unexplored.

The primary contribution of this paper is a control architecture for safe soft-landing on SCBs. Following spacecraft nominal control architectures, we assume a reference trajectory is provided by an offline algorithm. We integrate disturbance estimation, trajectory tracking of the reference trajectory, as well as safety and input constraint satisfaction throughout the maneuver. A nonlinear EHGO estimates gravitational uncertainties, while a feedback-linearizing, disturbance-canceling controller tracks the reference trajectory. State and input constraints are enforced via a CBF-based minimum-intervention layer. This architecture offers several advantages. The CBF-based layer eliminates the need for conservative planning, enabling the execution of aggressive trajectories. Furthermore, the framework is computationally efficient, making it suitable for autonomous spacecraft applications. Its effectiveness is demonstrated in simulation studies, which show consistent satisfaction of landing performance and safety constraints.

The remainder of this paper is organized as follows. Section \ref{sec:ProblemFormulation} outlines the problem formulation. Section \ref{sec:FeedbackControl} presents the disturbance observer and the feedback-linearizing control. Section \ref{sec:CBFControl} introduces the CBF-based constraints and formulates the safe control. Finally, Section \ref{sec:Simulations} provides numerical simulation results.

\section{Problem Formulation}\label{sec:ProblemFormulation}

Let $h : \BBR^n \rightarrow \BBR^l$ be continuously differentiable. Then, $h' : \BBR^n \rightarrow \BBR^{l \times n}$ is defined as $h'(x) \triangleq \frac{\partial h (x)}{\partial x}$. The Lie derivatives of $h$ along the vector field of $\psi:\BBR^n \rightarrow \BBR^{n \times l}$ is $L_{\psi}h \triangleq h'(x) \psi(x)$. Let $\| \cdot \|$ denote the 2-norm. The boundary of $\mathcal{A} \subseteq \BBR^n$ is denoted by $\text{bd }\mathcal{A}$. 

Let $\rho > 0$, and consider the soft-minimum function $\mathrm{softmin}_{\rho} : \BBR \times \cdots \times \BBR \to \BBR$ defined by
\begin{equation*}
    \mathrm{softmin}_\rho ( z_1, \ldots, z_N) \triangleq - \frac{1}{\rho} \mathrm{log} \sum_{i=1}^N e^{-\rho z_i},
\end{equation*}
which is a smooth lower bound of $\min \{ z_1, \ldots , z_N \}$.

\subsection{Equations of Motion}

Consider an irregularly shaped SCB (referred to as an asteroid) and a spacecraft with thrust capability in all orthogonal body directions. The equations of translational motion are
\begin{align}
    \dot{r}(t) &= v(t), \label{eq:vel}\\
    \dot{v}(t) &= - 2 \omega_{\times}v(t) - \omega_{\times}^2r(t) +  U'(r(t))^{\top} + \frac{u(t)}{m(t)}, \label{eq:accel}
\end{align}
where $r(t) \in \BBR^3$ is the position of the spacecraft's center of mass relative to the asteroid's center of mass, $v(t) \in \BBR^3$ is the velocity, $r(0) = r_0 \in \BBR^3$ and $v(0) = v_0 \in \BBR^3$ are the initial conditions, $m(t) >0$ is the spacecraft's mass, $u(t) \in \BBR^3$ is the thrust force, $\omega \in \BBR^3$ is the asteroid's angular velocity, $\omega_{\times} \in \BBR^{3 \times 3}$ is the skew-symmetric form of $\omega$, and $U: \mathbb{R}^3 \to \mathbb{R}$ is the asteroid's gravitational potential field. The spacecraft's mass is governed by
\begin{equation} \label{eq:mass_dynamics}
    \dot{m}(t) = - \alpha\|u(t)\|,
\end{equation}
where $m(0)>0$ is the initial mass and $\alpha >0$ is the mass depletion constant. The attitude of the spacecraft is controlled separately from its position.

\begin{figure}[t]
  \centering
  \includegraphics[width=0.7\columnwidth]{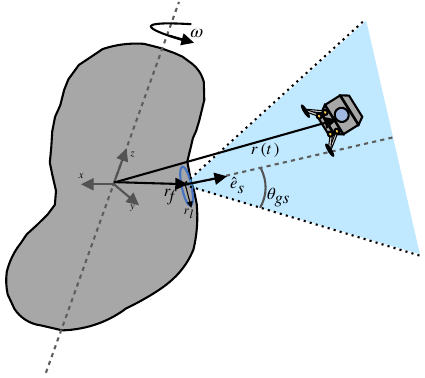}
  \vspace{-0.2 in}
  \caption{Spacecraft orbiting an irregularly shaped asteroid. The blue cone represents the position constraint \cref{eq:cons_pos}. Landing area is represented by the blue circle surrounding the desired landing position.}
  \label{fig:SpacecraftAsteroid}
  \vspace{-0.2 in}
\end{figure}

\subsection{Gravitational Field Model}\label{sec:Gravitational Model}

There are multiple sources of gravitational uncertainty, including uncertainty in the asteroid’s mass and mass distribution. We treat this uncertainty as perturbations to a nominal model. Consider the gravitational field model $U_{\rm m}: \mathbb{R}^3 \to \mathbb{R}$, which is known and 2-times continuously differentiable. The error is
\begin{equation}\label{eq:grav_disturbance}
    \Delta(r) \triangleq U(r) - U_{\rm m}(r).
\end{equation}
We make the following assumption:
\begin{enumA}
    \item $\frac{\partial \Delta}{\partial r}$ and $\frac{\partial^2 \Delta}{\partial r^2}$ are bounded. \label{ass:Bounded_Dist}
\end{enumA}

Assumption \ref{ass:Bounded_Dist} implies that model uncertainty induces a bounded error in acceleration. The point-mass Newtonian potential is one choice for $U_{\rm m}$. In cases where more information about the asteroid's geometry and composition is available, spherical harmonic models can be a more accurate representation \cite{scheeres2012orbital}.

\subsection{Asteroid Soft-Landing Problem}

The equations \cref{eq:vel,eq:accel,eq:mass_dynamics} can be written as
\begin{equation} \label{eq:System_Dynamics}
    \dot{x}(t) = f(x(t)) + g(x(t))u(t) + Dw(x(t)) + B \|u(t)\|,
\end{equation}
where $w(x) \triangleq \Delta'(r)^{\top}$ and
\begin{equation*}
    x \triangleq \begin{bmatrix} r \\ v \\ m \end{bmatrix}, \quad
    f(x) \triangleq 
    \begin{bmatrix}
        v\\
    - 2 \omega_{\times}v - \omega_{\times}^2r +  U_{\rm m}'(r)^{\top}\\
        0
    \end{bmatrix},
\end{equation*}
\begin{equation*}
    g(x) \triangleq 
    \begin{bmatrix}
        0_{3 \times 3} \\
        \frac{1}{m}I_{3\times3} \\
        0_{1 \times 3}
    \end{bmatrix}, \quad
    D \triangleq \begin{bmatrix}
                0_{3 \times 3} \\
                I_{3 \times 3} \\
                0_{1 \times 3}
            \end{bmatrix}, \quad
    B \triangleq \begin{bmatrix}
                0_{3 \times 1} \\
                0_{3 \times 1} \\
                -\alpha
            \end{bmatrix}.
\end{equation*}

The soft-landing maneuver is a controlled touchdown on the surface of the asteroid. The spacecraft must reach a final position $r_{\rm f} \in \BBR^3$ and velocity $v_{\rm f} \triangleq \omega_{\times} r_{\rm f}$ within pre-defined tolerances. Let $r_l>0$ be the allowable landing radius, $\delta>0$ be the allowable altitude offset, and $\varepsilon>0$ be the allowable speed error. Moreover, let $\hat{e}_s \in \BBR^3$ be the local perpendicular unit vector at the landing site. We formalize landing with the following definition.
\begin{definition} \rm{ \label{def:Landing}
    The spacecraft \cref{eq:vel,eq:accel,eq:mass_dynamics} \textit{lands} if there exists $t_l \in [0, \infty)$ such that $\|{r}(t_l) - {r_{\rm f}} \| \leq r_l$, $\|\hat{e}_s^{\top} ({r}(t_l) - r_{\rm f})\| \leq \delta$, and $\|{v}(t_l) - {v_{\rm f}} \| \leq \varepsilon$.
}
\end{definition}

In addition to landing, the spacecraft must satisfy a safety constraint. Consider the state-constraint function
\begin{equation}
    \psi(x) \triangleq \hat{e}_s^{\top} ({r} - \beta_{\rm gs}r_{\rm f}) - \| {r} - \beta_{\rm gs}r_{\rm f} \| \cos\theta_{\rm{gs}}, \label{eq:cons_pos}
\end{equation}
where $\theta_{gs} \in [0, \frac{\pi}{2}]$ is the landing glideslope angle and $\beta_{\rm gs} \triangleq 1 - \frac{r_l}{\|r_{\rm f}\| \text{atan} \theta_{gs}}$. The constraint $\psi(x) \geq 0$ is a glideslope constraint (see \cref{fig:SpacecraftAsteroid}), which forces the spacecraft to land inside the landing radius $r_l$. The set of allowable states is $\Psi \triangleq \{ x \in \BBR^7 : \psi(x) \geq 0\}$.

We also consider control constraints
\begin{equation}
    \phi_1(u) \triangleq T_{\max}^2 - \| u \|^2, \quad \phi_2(u) \triangleq \| u \|^2 - T_{\min}^2, \label{eq:Thrust_cons}
\end{equation}
where $T_{\rm{max}} > T_{\rm{min}} \geq 0$ are the limits on the magnitude of the thrust force. The set of admissible controls is $\Phi \triangleq \{ u \in \BBR^3 : \phi_1(u)\geq0, \phi_2(u) \geq 0 \}$.

The objective is to design a full-state feedback control such that: 
\begin{enumO}
    \item The spacecraft \cref{eq:vel,eq:accel,eq:mass_dynamics} lands. \label{Obj:landing}
    
    \item For all $t \in [0,t_l]$, $x(t) \in \Psi$ and $u(t) \in \Phi$. \label{Obj:safety}
\end{enumO}

We develop a control that tracks a reference trajectory. Let $x_{\rm r} : [0, \infty) \to \BBR^7$ and $u_{\rm r} : [0, \infty) \to \BBR^3$ be a reference state and control that satisfies the following conditions:
\begin{enumA}
    \setcounter{enumAi}{1}
    \item $(x,u) = (x_{\rm r}, u_{\rm r})$ satisfies \cref{eq:System_Dynamics} with $w = 0$, and the spacecraft lands. \label{ass:conv_traj}
    \item For all $t \in [0, \infty)$, $x_{\rm r}(t) \in \Psi$ and $u_{\rm r}(t) \in \Phi$.
    \item $u_{\rm r}$ is continuously differentiable on $[0, \infty)$. \label{assump:u_r_diff}
\end{enumA}

\begin{figure*}[t]
  \centering
  \includegraphics[width=\textwidth]{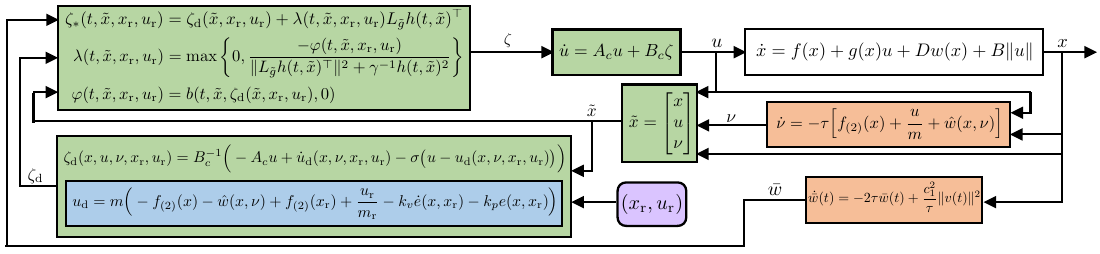}
  \vspace{-0.3 in}
  \caption{Closed-form safe control with state and input constraints under gravitational uncertainty. A desired control tracks a reference trajectory, while a safety-constrained control minimizes deviation from the desired input and guarantees safety despite gravitational uncertainty.}
  \label{fig:ControlArchitecture}
  \vspace{-0.15 in}
\end{figure*}

The control in this paper is structured in 3 stages (see \cref{fig:ControlArchitecture}). In the first stage, an EHGO is used to estimate $\Delta$. In the second stage, a feedback linearizing controller is used to track the reference trajectory. In the final stage, a control satisfying state and input constraints is generated from a minimum-intervention quadratic program.

\section{Feedback-Linearizing and Disturbance-Canceling Control} \label{sec:FeedbackControl}

This section presents the control developed to meet \ref{Obj:landing}.

\subsection{EHGO for Gravitational Uncertainty Estimation}

To compensate for gravitational uncertainty, we estimate $w$ using an EHGO similar to \cite{dunham_constrained_2016}. Let $\tau > 0$ be the gain, and consider the auxiliary variable $\nu: [0, \infty) \to \BBR^3$ that satisfies
\begin{equation} \label{eq:EHGO_Dynamics}
        \dot{\nu}(t) = -\tau \Big  [f_{(2)}(x(t)) + \frac{u(t)}{m(t)} + \hat{w}(x(t), \nu(t)) \Big],
\end{equation}
where $f_{(2)}$ is the function made up of rows 4 to 6 of $f$, $\hat{w}(x, \nu) \triangleq \tau v + \nu$ is the disturbance estimate, and $\nu(0) = -\tau v_0$. Define the estimation error $\tilde{w}(x, \nu) \triangleq w(x) - \hat{w}(x, \nu)$. The next result provides a computable upper bound on the estimation error. The proof is omitted for space.

\begin{fact}\label{lemma:Dist_est_conv}\rm
    Consider \cref{eq:System_Dynamics,eq:EHGO_Dynamics}, where \ref{ass:Bounded_Dist} is satisfied. Then, $\hat{w}$ and $\tilde{w}$ are bounded. Furthermore, for all $t \ge 0$,
    \begin{equation*}
        \|\tilde{w}(x(t), \nu(t))\| \le \sqrt{\bar{w}(t)},
    \end{equation*}
    where $\bar{w}: [0,\infty) \to \BBR$ satisfies
    \begin{equation}\label{eq:est_error_var_dyn}
        \dot{\bar{w}}(t) = -2\tau \bar{w}(t) + \frac{c_1^2}{\tau}\|v(t)\|^2,
    \end{equation}
    where $\bar{w}(0) \geq \|\tilde{w}(x(0), \nu(0))\|^2 \geq 0$ and $c_1>0$ is such that for all $r \in \BBR^3$, $c_1 \geq \|\frac{\partial^2 \Delta(r)}{\partial r^2}\|$.
\end{fact}


    


\subsection{Feedback Linearizing Tracking Control}

We now design a desired control that tracks the reference trajectory $(x_{\rm r}, u_{\rm r})$ and compensates for the uncertainty $w$. Let $x_{\rm r}$ be written as $x_{\rm r} = \begin{bmatrix} r_{\rm r}^{\top} & v_{\rm r}^\top & m_{\rm r} \end{bmatrix}$, where $r_{\rm r}(t) \in \BBR^3$, $v_{\rm r}(t) \in \BBR^3$ and $m_{\rm r}(t) \in \BBR$. Define $e \triangleq r- r_{\rm r}$. Let $k_p, k_v>0$, and consider the desired control
\begin{multline}\label{eq:desired_control}
    u_{\rm d}(x, \nu, x_{\rm r}, u_{\rm r}) \triangleq m \Big( -f_{(2)}(x) - \hat{w}(x,\nu) \\
    + f_{(2)}(x_{\rm r}) + \frac{u_{\rm r}}{m_{\rm r}}- k_v\dot{e} - k_pe \Big).
\end{multline}

The following result provides convergence properties for $e$ and $\dot{e}$. The proof is omitted for space.

\begin{lemma}\rm\label{lemma:control_conv}
    Consider \cref{eq:System_Dynamics} and \cref{eq:EHGO_Dynamics}, where $u = u_{\rm d}$ and \ref{ass:Bounded_Dist} is satisfied. Then, there exists $\eta > 0$ such that
    \begin{equation*}
        \left \| \begin{bmatrix} e(t) \\ \dot{e}(t) \end{bmatrix} \right \| \to \left [ 0, \frac{\eta}{\tau} \right ] \quad \text{as} \quad t \to \infty.
    \end{equation*}
\end{lemma}

\Cref{lemma:control_conv} and \ref{ass:conv_traj} imply that for a sufficiently large $\tau$, \ref{Obj:landing} is satisfied. However, the control $u_{\rm d}$ does not necessarily satisfy \ref{Obj:safety}.

\section{CBFs for State and Input Constraints} \label{sec:CBFControl}

This section presents a minimum-intervention control to satisfy \ref{Obj:safety}. Define
\begin{equation*}
    \mathcal{S}_{\rm s} \triangleq \{ (x,u) \in \BBR^{7} \times \BBR^{3} : x \in \Psi \text{ and } u \in \Phi \}.
\end{equation*}
If $(x(t), u(t)) \in \mathcal{S}_{\rm s}$ for all $t \in [0, \infty)$, then \ref{Obj:safety} is satisfied.

\subsection{Control Dynamics and Soft-Minimum Relaxed CBF}

Since $\phi_1$ and $\phi_2$ in \cref{eq:Thrust_cons} depend explicitly on $u$, they are not valid candidate CBFs. We adopt the method in \cite{rabiee_closed-form_2024,rabiee_composition_2024} to transform these input constraints into controller-state constraints by introducing control dynamics. Let $u$ satisfy
\begin{equation}\label{eq:control_dynamics}
    \dot{u}(t) = A_c u(t) + B_c\zeta(t),
\end{equation}
where $A_c \in \BBR^{3 \times 3}$ is asymptotically stable, $B_c \in \BBR^{3 \times 3}$ is nonsingular, $u(0) = u_{\rm r}(0)$ is the initial condition,  and $\zeta: [0,\infty) \to \BBR^3$ is the surrogate control, which is the input to \cref{eq:control_dynamics}. Next, consider the desired surrogate control $\zeta_{\rm d}: \BBR^7 \times \BBR^3 \times \BBR^3 \times \BBR^7 \times \BBR^3 \to \BBR^3$ defined by
\begin{multline} \label{eq:desired_surrogate}
    \zeta_{\rm d}(x,u, \nu, x_{\rm r}, u_{\rm r}) \triangleq B_c^{-1}\Big(-A_c u + \dot{u}_{\rm d}(x, \nu,x_{\rm r}, u_{\rm r}) \\- \sigma \big(u - u_{\rm d}(x, \nu,x_{\rm r}, u_{\rm r}) \big)\Big),
\end{multline}
where $\sigma > 0$. Proposition 8 in \cite{rabiee_closed-form_2024} shows that if $\zeta = \zeta_{\rm d}$, then $u$ converges exponentially to $u_{\rm d}$.

Since the control $u$ is a state of the control \cref{eq:control_dynamics}, the input constraints are now state constraints of the cascade \cref{eq:System_Dynamics,eq:EHGO_Dynamics,eq:control_dynamics}. Note that $\phi_1$ and $\phi_2$ have relative degree one with respect to the cascade with input $\zeta$. However, $\psi$ has relative degree 3 with respect to the cascade with input $\zeta$. Thus, we adopt a higher-order approach to develop a candidate CBF.

Let $\beta,\beta_1: \BBR \to \BBR$ be extended class-$\mathcal{K}$ functions that are 3-times and 2-times continuously differentiable. Then, consider $\psi_1:\BBR^7 \to \BBR$ and $\psi_2:\BBR^7 \times \BBR^3 \to \BBR$ defined by
\begin{align}
    \psi_1(x) &\triangleq L_f \psi(x) + \beta(\psi(x)), \notag\\
    \psi_2(x,u) &\triangleq L_f \psi_1(x) + L_g \psi_1(x) u \notag\\
    &\qquad + L_D \psi_1(x) w(x) + \beta_{1}(\psi_1(x)). \label{eq:pos_valid_CBF}
\end{align}
Note that \cref{eq:pos_valid_CBF} depends on the unknown function $w$. Thus, we consider $\ubar{\psi}_2: [0,\infty) \times \BBR^7 \times \BBR^3 \times \BBR^3 \to \BBR$ given by
\begin{multline}\label{eq:comp_gs_HOCBF}
    \ubar{\psi}_2(t,x,u,\nu) \triangleq L_f \psi_1(x) + L_g \psi_1(x) u + L_D \psi_1(x) \hat{w}(x,\nu) \\
    - \sqrt{\varepsilon + \|L_D \psi_1(x)^\top\|^2} \sqrt{\bar{w}(t)} + \beta_{1}(\psi_1(x)), 
\end{multline}
where $\varepsilon>0$. The next result shows that \cref{eq:comp_gs_HOCBF} is a computable lower bound of \cref{eq:pos_valid_CBF}. The proof is omitted for space.

\begin{proposition}
    Consider \cref{eq:System_Dynamics,eq:EHGO_Dynamics,eq:est_error_var_dyn} where \ref{ass:Bounded_Dist} is satisfied. Let $u \in \BBR^3$. Then, for all $t\geq0$, $\ubar{\psi}_2(t,x(t),u,\nu(t)) \leq \psi_2(x(t), u)$. 
\end{proposition}

The cascade of \cref{eq:System_Dynamics,eq:EHGO_Dynamics,eq:control_dynamics} is
\begin{equation}\label{eq:Augmented_Dynamics}
    \dot{\tilde{x}} = \tilde{f}(\tilde{x})+\tilde{g}(\tilde{x})\zeta + \tilde{D}w(x)
\end{equation}
where
\begin{align*}
    \tilde{x} &\triangleq \begin{bmatrix}
        x \\
        u \\
        \nu
    \end{bmatrix}, \quad
    \tilde{f}(\tilde{x}) \triangleq \begin{bmatrix}
        f(x) + g(x)u + B\|u\| \\
        A_c u \\
        -\tau \Big  [f_{(2)}(x) + \frac{u}{m} + \hat{w}(x, \nu) \Big]
    \end{bmatrix},\\
    \tilde{g}(\tilde{x}) &\triangleq \begin{bmatrix} 0_{7 \times 3} \\ B_c \\ 0_{3 \times 3} \end{bmatrix}, \quad \tilde{D} \triangleq\begin{bmatrix}D \\ 0_{3 \times 3} \\ 0_{3 \times 3} \end{bmatrix}.
\end{align*}
Next, define
\begin{align*}
    \bar{\mathcal{S}}(t) &\triangleq \{\tilde{x} \in \BBR^{13}: \psi(x) \geq 0, \psi_1(x) \geq 0, \ubar{\psi}_2(t,x,u,\nu) \geq 0, \\
    &\qquad \phi_1(u) \geq 0, \phi_2(u) \geq 0\},\\
    \bar{\mathcal{H}}(t) &\triangleq \{ \tilde{x} \in \BBR^{13}: \ubar{\psi}_2(t,x,u,\nu)\geq0, \phi_1(u)\geq0, \phi_2(u)\geq0 \}.
\end{align*}
If $\tilde{x}(0) \in \bar{\mathcal{S}}(0)$ and for all $t \in [0, \infty)$, $\tilde{x}(t) \in \bar{\mathcal{H}}(t)$, then for all $t \in [0, \infty)$, $\tilde{x}(t) \in \bar{\mathcal{S}}(t) \subset \mathcal{S}_{\rm s} \times \BBR^3$ \cite[Theorem 3]{tan_high-order_2022}. This motivates us to consider a candidate CBF whose zero-superlevel set is a subset of $\bar{\mathcal{H}}(t)$.

We use a composite CBF \cite{rabiee_closed-form_2024}. Let $\rho > 0$ and consider $h: [0,\infty) \times \BBR^{13} \to \BBR$ given by
\begin{equation*}
    h(t,\tilde{x}) \triangleq \mathrm{softmin}_\rho \Big (k_{\rm{gs}}\ubar{\psi}_2(t,x,u,\nu), k_{\rm{u}}\phi_1(u), k_{\rm{u}}\phi_2(u)  \Big ).
\end{equation*}
where $k_{\rm{gs}}, k_{\rm{u}} > 0$ are used to normalize magnitudes. In the case where $T_{\min} = 0$, $\phi_2$ is not included in $h$. The zero-superlevel set of $h$ is
\begin{equation*}
    \mathcal{H}(t) \triangleq \{ \tilde{x} \in \BBR^{13}: h(t,\tilde{x}) \geq 0 \},
\end{equation*}
and \cite[Proposition 2]{safari_safe_2025} implies that $\mathcal{H}(t) \subset \bar{\mathcal{H}}(t)$ and as $\rho \to \infty, \mathcal{H}(t) \to \bar{\mathcal{H}}(t)$. Next, consider the set
\begin{align*}
        \mathcal{B}(t) &\triangleq \{ \tilde{x} \in \text{bd } \mathcal{H}(t) : \frac{\partial h(t,\tilde{x})}{\partial t} + L_{\tilde{f}}h(t,\tilde{x}) \\
        &\qquad + L_{\tilde{D}}h(t,\tilde{x})w(x) \leq 0\},
\end{align*}
which is the set of all augmented states on the boundary of $\mathcal{H}(t)$ where the time derivative of $h$ along the uncontrolled flow of \cref{eq:Augmented_Dynamics} is non-positive. We assume that $\zeta$ directly influences the time derivative of $h$ on $\mathcal{B}(t)$, specifically:
\begin{enumA}
    \setcounter{enumAi}{4}
    \item For all $(t,\tilde{x}) \in [0,\infty) \times \mathcal{B}(t)$, $L_{\tilde{g}}h(t,\tilde{x}) \neq 0$. \label{ass:cbf}
\end{enumA}
The following result shows that $h$ is a relaxed control barrier function (R-CBF). The proof is similar to \cite[Proposition 8]{safari_time-varying_2025}.

\begin{proposition}\rm \label{prop:valid_CBF}
    Assume \ref{ass:cbf} is satisfied. Then, for all $(t,\tilde{x}) \in [0,\infty) \times \mathcal{H}(t)$,
    \begin{multline*}
        \sup_{\zeta \in \BBR^3} \Big[\frac{\partial h(t,\tilde{x})}{\partial t} + L_{\tilde{f}}h(t,\tilde{x}) + L_{\tilde{g}}h(t,\tilde{x})\zeta \\
        + L_{\tilde{D}}h(t,\tilde{x})w(x)\Big] \geq 0.
    \end{multline*}
\end{proposition}

Since $h$ is a R-CBF on $\mathcal{H}(t)$, it is possible to construct a control such that $\mathcal{S}(t) \triangleq \bar{\mathcal{S}}(t) \cap \mathcal{H}(t)$ is forward invariant \cite[Prop. 9]{safari_time-varying_2025}.

\subsection{Optimal and Safe Control}

We use the R-CBF $h$ to design a control $\zeta$ that enforces forward invariance of $\mathcal{S}$ while minimizing the cost function
\begin{equation} \label{eq:Cost_function}
    \mathcal{J}(\tilde{x}, x_{\rm r}, u_{\rm r}, \hat{\zeta}, \hat{\kappa}) \triangleq \| \hat{\zeta} - \zeta_{\rm d}(\tilde{x},x_{\rm r}, u_{\rm r}) \|^2 + \gamma \hat{\kappa}^2,
\end{equation}
where $\hat{\zeta}$ is the optimization variable, $\hat{\kappa}$ is a slack variable, and $\gamma>0$. Let $\alpha: \BBR \to \BBR$ be an extended class-$\mathcal{K}$ function, and consider the constraint function $b: [0,\infty) \times \BBR^{13} \times \BBR^3 \times \BBR \to \BBR$ defined as
\begin{multline} \label{eq:safety_cons}
    \ubar{b}(t,\tilde{x}, \hat{\zeta}, \hat{\kappa}) \triangleq 
    \frac{\partial h(t,\tilde{x})}{\partial t} 
    + L_{\tilde{f}} h(t,\tilde{x}) + L_{\tilde{g}} h(t,\tilde{x}) \hat{\zeta} \\
    + L_{\tilde{D}} h(t,\tilde{x})\, \hat{w}(x,\nu) 
    - \sqrt{\varepsilon + \|L_{\tilde{D}} h(t,\tilde{x})^\top\|^2}\, \sqrt{\bar{w}(t)}  \\
    + \alpha\big(h(t,\tilde{x})\big) 
    + \hat{\kappa} h(t,\tilde{x}),
\end{multline}
and note that $b$ is a computable lower bound of $\dot h(t,x) + \alpha(h(t,x))+\kappa h(t,x)$ along the trajectory of \cref{eq:Augmented_Dynamics}.

For each $(t,\tilde{x}) \in [0,\infty) \times \BBR^{13}$, the minimizer of \cref{eq:Cost_function} subject to $b(t,\tilde{x}, \hat{\zeta}, \hat{\kappa}) \geq 0$ can be derived from the first-order necessary conditions for optimality \cite{rabiee_closed-form_2024}, and is given by
\begin{equation}\label{eq:safe_control}
    \zeta_*(t, \tilde{x}, x_{\rm r}, u_{\rm r}) \triangleq \zeta_{\rm d}(\tilde{x},x_{\rm r}, u_{\rm r}) +\lambda(t, \tilde{x}, x_{\rm r}, u_{\rm r})L_{\tilde{g}}h(t,\tilde{x})^\top,
\end{equation}
\begin{equation}\label{eq:safe_kappa}
    \kappa_*(t, \tilde{x}, x_{\rm r}, u_{\rm r}) \triangleq \gamma^{-1}h(t,\tilde{x})\lambda(t, \tilde{x}, x_{\rm r}, u_{\rm r}),
\end{equation}
where $\lambda,\varphi: [0,\infty) \times \BBR^{13} \times \BBR^7 \times \BBR^3 \to \BBR$ are defined by
\begin{equation}\label{eq:lambda_safety}
\lambda(t,\tilde{x},x_{\rm r},u_{\rm r}) \triangleq \max \left\{0, \frac{-      \varphi(t,\tilde{x},x_{\rm r},u_{\rm r})}{\|L_{\tilde g}h(t,\tilde{x})^\top\|^2 +\gamma^{-1}h(t,\tilde{x})^2} \right \},
\end{equation}
\begin{equation} \label{eq:safe_omega}
    \varphi(t, \tilde{x}, x_{\rm r}, u_{\rm r}) \triangleq b(t,\tilde{x}, \zeta_{\rm d}(\tilde{x}, x_{\rm r}, u_{\rm r}),0).
\end{equation}
The parameter $\gamma$ weights $\kappa$ in the cost \eqref{eq:Cost_function} and influences control aggressiveness. The following theorem is the main result on constraint satisfaction. The proof is in \cite[Theorem 2]{safari_safe_2025}.

\begin{theorem}\rm{ \label{thm:safety}
    Consider \cref{eq:System_Dynamics,eq:EHGO_Dynamics,eq:est_error_var_dyn}, where \ref{ass:Bounded_Dist} and \ref{ass:cbf} are satisfied. Let $u$ satisfy \cref{eq:control_dynamics}, where $\zeta = \zeta_*$ and $\zeta_*$ is given by \cref{eq:safe_control,eq:lambda_safety,eq:safe_omega}. Assume that $h'$ is locally Lipschitz in $\tilde{x}$. Then, for all $\tilde{x}(0) \in \mathcal{S}(0)$, the following statements hold:
    \begin{enumerate}[label=\alph*)]
        \item There exists a maximum value $t_{\rm m}\in(0,\infty]$ such that \cref{eq:Augmented_Dynamics} with $\zeta=\zeta_*$ has a unique solution on $[0,t_{\rm m})$.
        \item For all $t \in [0,t_{\rm m})$, $\tilde{x}(t) \in \mathcal{S}(t) \subseteq \mathcal{S}_{\rm s} \times \BBR^3$.
    \end{enumerate}
}
\end{theorem}

\Cref{thm:safety} shows that $(x(t),u(t)) \in \mathcal{S}_{\rm s}$, thus, achieving \ref{Obj:safety} while producing a control $u$ that approximates $u_{\rm d}$, which is designed to achieve \ref{Obj:landing}.

\section{Simulation Results} \label{sec:Simulations}

We present simulations, where $U(r)$ is the spherical harmonics model from  \cite{arenasuribe2026higherordergravitationalmodelstutorial}. The known model $U_{\rm m}$ is the point-mass Newtonian model with mass uncertainty of 25\%. Reference trajectories $(x_{\rm r}, u_{\rm r})$ are obtained with a trajectory generation algorithm \cite{malyuta_convex_2022}. We consider a spacecraft with $m(0) = 700 \text{ kg}$, $\alpha = 4.53 \times 10^{-4} \frac{\rm s}{\rm m}$, $T_{\min} = 0 \text{ N}$ and $T_{\max} = 30 \text{ N}$. The landing tolerances are $r_l = 1.5 \,\text{m}$, $\delta = 1.0 \,\text{m}$, and $\varepsilon = 0.5 \,\frac{\text{m}}{\text{s}}$. The constraint function \cref{eq:cons_pos} uses $\theta_{\rm gs} = \frac{\pi}{4}$. We implement the control architecture at 25 Hz. For details on the control parameters and implementation see the repository.\footnote{\url{https://github.com/FelipeArenasUribe/SCB_Position_Safe_Landing}}   

For the following examples, executing the trajectory with $u = u_{\rm r}$ results in position constraint violations; using $u = u_{\rm d}$ alone results in input constraint violations; and $u = u_{\rm d}$ with saturation results in position constraint violations. These outcomes expose the limitations of direct trajectory execution and feedback control without explicitly addressing constraints.

\subsection{Ellipsoidal Asteroid with Uniform Density}

We consider an ellipsoidal asteroid with semi-axes $a = 1000 \, \text{m}$, $b = c = 400 \, \text{m}$ and density $\rho = 1380 \frac{\rm{kg}}{\rm m^3}$, which is typical of C-type asteroids. The asteroid rotates about its principal axis with the largest moment of inertia with a synodic period of $4 \, \text{h}$. The initial and desired states are $r_0 = \begin{bmatrix} 1927.2 & -374.6 & -954 \end{bmatrix}^{\top} \rm m$, $r_{\rm f} = \begin{bmatrix} 400 & 0 & 0 \end{bmatrix}^{\top} \rm m$ and $v_0 = \begin{bmatrix} -1.64 & -3.02 & -3.64 \end{bmatrix}^{\top} \frac{\rm m}{\rm s}$.

\begin{figure}[ht]
  \centering
  \includegraphics[width=0.75\columnwidth]{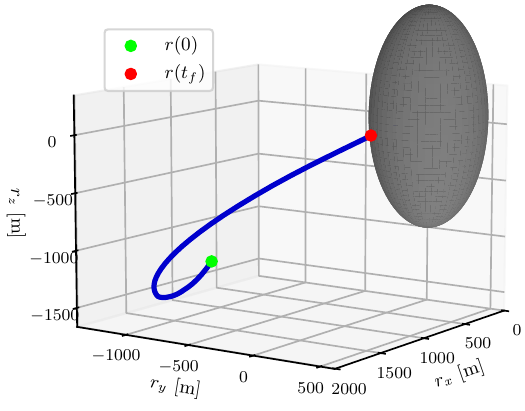}
  \vspace{-0.18 in}
  \caption{Spacecraft trajectory during the landing maneuver on the ellipsoidal asteroid.}
  \label{fig:Ellipsoid_Trajectory}
  \vspace{-0.15 in}
\end{figure}

Landing is achieved at $t_l = 840 \text{ s}$, and $\| r(t_l) -r_{\rm f} \| = 1.27 \text{ m}$, $\hat{e}^{\top}(r(t_l) -r_{\rm f}) = 0.982 \text{ m}$ and $\| v(t_l) - v_{\rm f} \| = 0.262 \frac{\rm m}{\rm s}$. Figure \ref{fig:Ellipsoid_Trajectory} shows the trajectory of the spacecraft during the maneuver. 

\begin{figure}[ht]
    \centering
    \includegraphics[width=0.8\columnwidth]{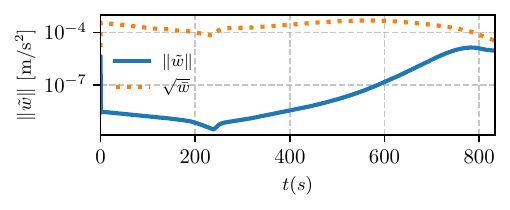}
    \vspace{-0.2 in}
    \caption{Disturbance estimate error for ellipsoidal asteroid case. The error is upper bounded by the computable parameter $\sqrt{\bar{w}}$.}
    \label{fig:Ellipsoid_disturbance}
    \vspace{-0.3 in}
\end{figure}

The EHGO estimates the gravitational uncertainty and $\sqrt{\bar{w}} \geq \|\tilde{w}\|$ as illustrated in \cref{fig:Ellipsoid_disturbance}. The magnitude of the error increases as the spacecraft approaches the asteroid due to the increasing influence of higher-order effects.

\begin{figure}[H]
  \centering
  \includegraphics[width=0.8\columnwidth]{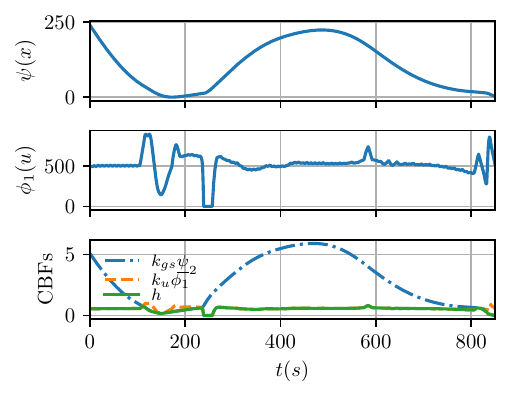}
  \vspace{-0.22 in}
  \caption{State and input constraints remain positive, ensuring safe execution of the soft-landing maneuver.} 
  \label{fig:Ellipsoid_cbfs}
  \vspace{-0.35 in}
\end{figure}

\begin{figure}[ht]
  \centering
  \includegraphics[width=0.8\columnwidth]{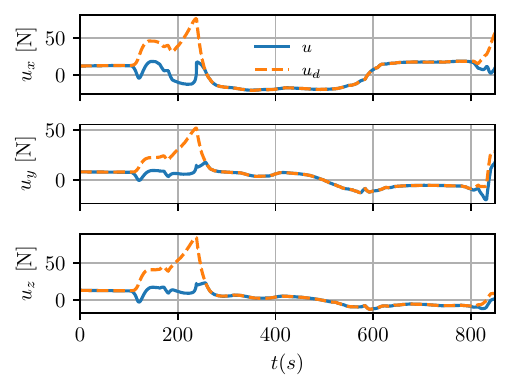}
  \vspace{-0.2 in}
  \caption{Desired control $u_{\rm d}$ and control $u$ during the maneuver. The controller modifies the input in real time to ensure constraint satisfaction.}
  \label{fig:Ellipsoid_inputs}
  \vspace{-0.2 in}
\end{figure}

Figure \ref{fig:Ellipsoid_cbfs} shows that state and input constraints are satisfied. As shown in Figure \ref{fig:Ellipsoid_inputs}, the controller actively modifies the desired control input at two periods of time, in both cases, the unmodified control would have violated the position constraint.

\subsection{Asteroid with Non-uniform Density}

We consider an irregularly shaped asteroid with non-uniform density, modeled as the union of two ellipsoids with uniform densities. The primary has semi-axes $a=1000\text{ m}, b=c=400\text{ m}$ and density $\rho_1 = 1460 \frac{\rm{kg}}{\rm m^3}$, while the secondary has $a=800\text{ m}, b=c=400\text{ m}$ and $\rho_2 = 1380 \frac{\rm{kg}}{\rm m^3}$. The asteroid rotates about its principal axis with the largest moment of inertia with a synodic period of $4\text{ h}$. The initial and desired states are $r_0 = [1927.2,-374.6,-954]^{\top} \rm m$, $r_{\rm f} = [600,0,-253.5]^{\top} \rm m$ and $v_0 = [-1.64,-3.02,-3.64]^{\top} \frac{\rm m}{\rm s}$.

\begin{figure}[ht]
  \centering
  \includegraphics[width=0.75\columnwidth]{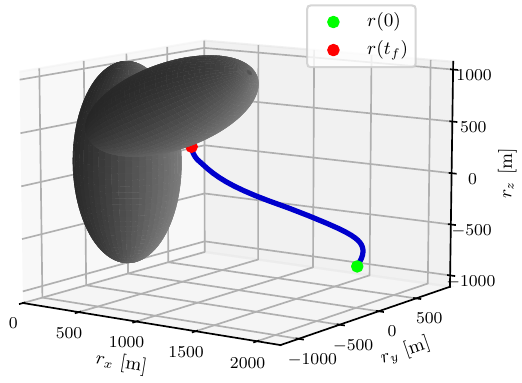}
  \vspace{-0.1 in}
  \caption{Spacecraft trajectory during the landing maneuver on the irregularly shaped asteroid.}
  \label{fig:Irregular_trajectories}
  \vspace{-0.28 in}
\end{figure}

Figure~\ref{fig:Irregular_trajectories} shows the trajectory of the spacecraft using the proposed control. Landing is achieved at $621 \text{ s}$, with performance errors $\| r(t_l) -r_{\rm f} \| = 1.476 \text{ m}$, $\hat{e}^{\top}(r(t_l) -r_{\rm f}) = 0.33 \text{ m}$ and $\| v(t_l) - v_{\rm f} \| = 0.203 \frac{\rm m}{\rm s}$.

\begin{figure}[H]
  \centering
  \includegraphics[width=0.8\columnwidth]{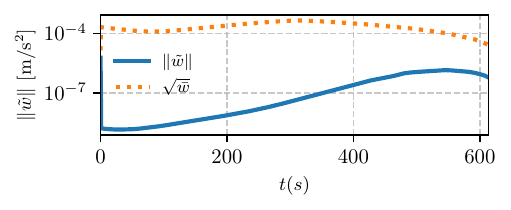}
  \vspace{-0.2 in}
  \caption{Disturbance estimate error for irregularly shaped asteroid.}
  \label{fig:Irregular_dist}
  \vspace{-0.3 in}
\end{figure}

\begin{figure}[ht]
  \centering
  \includegraphics[width=0.8\columnwidth]{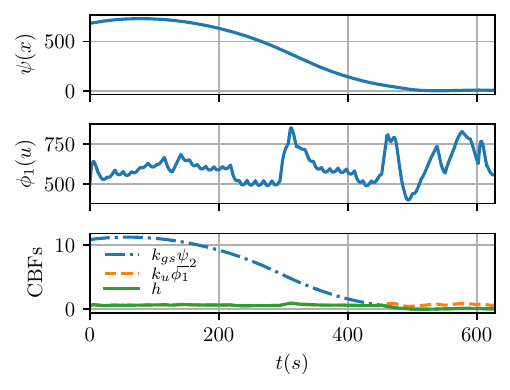}
  \vspace{-0.2 in}
  \caption{State and input constraints, and composite CBF with the corresponding higher-order CBFs.}
  \label{fig:Irregular_cbfs}
  \vspace{-0.3 in}
\end{figure}

\begin{figure}[ht]
  \centering
  \includegraphics[width=0.8\columnwidth]{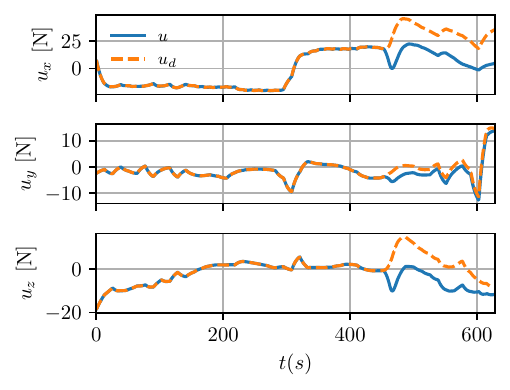}
  \vspace{-0.2 in}
  \caption{Desired input $u_{\rm d}$ and control $u$ during the landing maneuver on the irregularly shaped asteroid.}
  \label{fig:Irregular_inputs}
\end{figure}

The disturbance estimation error and its upper bound are shown in \cref{fig:Irregular_dist}. Figure~\ref{fig:Irregular_cbfs} shows input and state constraints are satisfied. Figure~\ref{fig:Irregular_inputs} shows that the control input is modified during the final 150 seconds to satisfy the position constraint.

\bibliographystyle{IEEEtran}
\bibliography{References}

\end{document}